# On certain systems of $4$ Ordinary Differential Equations $dx_n(t)/dt = P_2^{(n)}[x_m(t)]/[x_1(t)+x_2(t)]$, $n,m=1,2,3,4$, where $P_2^{(n)}(x_m)$ are $4$ polynomials of degree $2$ in the $4$ variables $x_m$

**Francesco Calogero**
Physics Department, University of Rome "La Sapienza", Rome, Italy
Istituto Nazionale di Fisica Nucleare, Sezione di Roma 1, Rome, Italy
Istituto Nazionale di Alta Matematica, Gruppo Nazionale di Fisica Matematica, Italy

## Abstract

Some interesting (periodic!) solutions of certain systems of 4 *nonlinear* Ordinary Differential Equations

$$dx_n(t)/dt = P_2^{(n)}[x_m(t)]/[x_1(t)+x_2(t)], \ n,m=1,2,3,4,$$

where $P_2^{(n)}[x_m(t)]$ are 4 polynomials of degree 2 in the 4 dependent variables $x_m(t)$, are obtained.

## 1. Introduction

It was quite recently pointed out [1] that the system of 4 *nonlinear* Ordinary Differential Equations (ODEs)

$$\dot{x}_n(t) = P_2^{(n)}[x_m(t)]/[x_1(t)+x_2(t)], \ n,m=1,2,3,4, \qquad (1)$$

is relevant to investigate an agricultural disease quite important for the citrus industry in China.

**Notation**: above and hereafter $n$ and $m$ (and, see below, as well $m_1$ and $m_2$) are 4 indices each taking the 4 values $1,2,3,4$, while (see below) $\nu$, $\ell$, $\ell_1$, $\ell_2$, $\mu$ are 5 indices taking respectively the 3 values $\nu = 1,3,4$, $\ell = 0,1,2$, $\ell_1 = 0,1,2$, $\ell_2 = 0,1,2$, $\mu = 1,2,3$, and instead the 3 indices $s$, $p$ and $j$ are 3 indices taking respectively the 2 values $s=1,2$, the 6 values $p=1,2,...,6$, and the 10 values $j=1,2,...,10$; while $P_2^{(n)}(x_m)$ are 4 polynomials of degree 2 in the 4 dependent variables $x_m$,

$$P_2^{(n)}(x_m) = a^{(n)} + \sum_{m=1}^{4}\left[b_m^{(n)}x_m + c_m^{(n)}(x_m)^2\right] + \sum_{m_1>m_2=1}^{4}\left[d_{m_1 m_2}^{(n)}x_{m_1}x_{m_2}\right] ; \qquad (2)$$

each of these 4 polynomials features $1+2\cdot 4 + 6 = 15$ coefficients $a^{(n)}$, $b_m^{(n)}$, $c_m^{(n)}$, $d_{m_1 m_2}^{(n)}$ (with $m_1 > m_2$), hence altogether 60 *a priori* (of course $t$-independent) *arbitrary* coefficients; the *independent* variable $t$ is, as usual, hereafter considered to represent *time*; and, as usual, we shall hereafter denote by a *superimposed dot* a differentiation with respect to $t$. ∎

In this paper we introduce 2 simple *Ansatzen* for the solutions $x_n(t)$ of the system of ODEs (1)—the second of which implies that they are *all* very simple *periodic* functions of $t$ with the *same* period $T = 2\pi/\omega$ (see below)—and identify the restrictions they imply on the 60 coefficients $a^{(n)}$, $b_m^{(n)}$, $c_m^{(n)}$, $d_{m_1 m_2}^{(n)}$ (with $m_1 > m_2$). These findings are reported in the following 3 **Sections**, complemented by the 3 **Appendices A**, **B** and **C**. Summarizing remarks are provided in the concluding **Section 4**.

## 2. Ansatz1

**Ansatz1** reads as follows:

$$x_1(t) = x_1(0) + y_0(t) , \quad x_2(t) = x_2(0) - y_0(t) , \tag{3a}$$

$$x_3(t) = x_3(0) + y_1(t) , \quad x_4(t) = x_4(0) + y_2(t) ; \tag{3b}$$

obviously implying

$$\dot{x}_1(t) = \dot{y}_0(t) , \quad \dot{x}_2(t) = -\dot{y}_0(t) , \quad \dot{x}_3(t) = \dot{y}_1(t) , \quad \dot{x}_4(t) = \dot{y}_2(t) ; \tag{3c}$$

hence it introduces the 4 *initial values* $x_n(0)$ of the 4 dependent variables $x_n(t)$ and (only!) 3 new variables $y_\ell(t)$, themselves now clearly required to satisfy the 3 *initial conditions*

$$y_\ell(0) = 0 . \tag{3d}$$

Note that this **Ansatz1** implies moreover (see (3a)) that

$$x_1(t) + x_2(t) = x_1(0) + x_2(0) \tag{3e}$$

is $t$-independent:

$$\dot{x}_1(t) + \dot{x}_2(t) = 0 ; \tag{3f}$$

hence it requires (see eqs. (1) and (2)) that

$$P_2^{(1)}(x_m) + P_2^{(2)}(x_m) = 0 , \tag{3g}$$

implying the following 15 *simple constraints* on the $2(1 + 2 \cdot 4 + 3 \cdot 4/2) = 30$ coefficients $a^{(s)}$, $b_m^{(s)}$, $c_m^{(s)}$, $d_{m_1 m_2}^{(s)}$ (with $m_1 > m_2$) of the 2 polynomials $P_2^{(s)}(x_m)$:

$$\begin{aligned} a^{(1)} &= -a^{(2)}, \quad b_m^{(1)} = -b_m^{(2)}, \quad c_m^{(1)} = -c_m^{(2)} , \\ d_{m_1 m_2}^{(1)} &= -d_{m_1 m_2}^{(2)} , \quad m_1 > m_2 , \end{aligned} \tag{4}$$

reducing them to, say, the only 15 coefficients $a^{(1)}$, $b_m^{(1)}$, $c_m^{(1)}$, $d_{m_1 m_2}^{(1)}$ (with $m_1 > m_2$); and it moreover reduces the number of actually evolving variables from the 4 variables $x_n(t)$ to the 3 variables $y_\ell(t)$, requiring these new variables to satisfy the system of 3 *nonlinear* (but now only *quadratic*) ODEs

$$\dot{y}_\ell(t) = Q_2^{(\ell)}[y_{\ell'}(t)], \quad \ell, \ell' = 0, 1, 2 , \tag{5a}$$

featuring now only 3 dependent variables $y_\ell(t)$ and the 3 *second-degree* polynomials

$$Q_2^{(\ell)}(y_0, y_1, y_2) = \alpha^{(\ell)} + \sum_{\ell_1=0}^{2} \left[ \beta_{\ell_1}^{(\ell)} y_{\ell_1} + \gamma_{\ell_1}^{(\ell)} (y_{\ell_1})^2 \right] + \sum_{\ell_1 > \ell_2 = 0}^{2} \left[ \eta_{\ell_1 \ell_2}^{(\ell)} y_{\ell_1} y_{\ell_2} \right] . \tag{5b}$$

These 3 polynomials $Q_2^{(\ell)}(y_0, y_1, y_2)$ feature now altogether $3+2\cdot3\cdot3+3\cdot6/2 = 30$ coefficients $\alpha^{(\ell)}, \beta_{\ell_1}^{(\ell)}, \gamma_{\ell_1}^{(\ell)}, \eta_{\ell_1\ell_2}^{(\ell)}$, which can be easily expressed in terms of the 4 *initial data* $x_n(0)$ and the original 60 coefficients $a^{(n)}, b_m^{(n)}, c_m^{(n)}, d_{m_1m_2}^{(n)}$ (see (2); themselves now required to satisfy the 15 *simple constraints* (4), and being therefore reduced to only 45). Relevant formulas implied by this **Ansatz1** are reported in **Appendix A**.

### 3. Ansatz 2

This second **Ansatz2** (consistent with (3d)) reads as follows:

$$y_\ell(t) = f_{1+2\ell}\sin(\omega t) + f_{2+2\ell}[1 - \cos(\omega t)] \ . \tag{6}$$

There are in this **Ansatz2** only 7 *a priori arbitrary* parameters: $\omega$ and the 6 parameters $f_p, p = 1, 2, ..., 6$. We hereafter assume them to be *all real* numbers, and (for definiteness), $\omega$ to be *positive*, $\omega > 0$.

It is then easily seen that this **Ansatz2** provides a *solution* of the *nonlinear* system of 3 ODEs (5a) with (5b), if there hold 15 *algebraic* equations relating the 7 quantities $\omega$ and $f_p$ to the 30 coefficients $\alpha^{(\ell)}, \beta_{\ell_1}^{(\ell)}, \gamma_{\ell_1}^{(\ell)}, \eta_{\ell_1\ell_2}^{(\ell)}$ (as shown in **Appendix B**).

### 4. Summarizing remarks

We now assume that the reader has digested the results presented above, as well as those provided in the 3 **Appendices A**, **B** and **C**, and is ready for a final summary of the findings reported in this paper.

In this paper we focussed on the system of 4 *nonlinearly-coupled* ODEs (1) with (2), featuring the 60 (*t*-independent) coefficients $a^{(n)}, b_m^{(n)}, c_m^{(n)}, d_{m_1m_2}^{(n)}$ (with $m_1 > m_2$).

When investigating such dynamical systems, 2 *alternative* approaches may be employed. One might consider the 60 coefficients $a^{(n)}, b_m^{(n)}, c_m^{(n)}, d_{m_1m_2}^{(n)}$ characterizing the system of 4 ODEs (1) with (2) as *given input* data, and investigate the *t*-evolution of the solutions of the system of ODEs (say, for any assignment of initial data $x_m(0)$). One might instead make *a priori* the assumption that those solutions feature a certain behavior—for instance, to be *completely periodic* with a given period $T$—and focus on the *restrictions* on the 60 coefficients $a^{(n)}, b_m^{(n)}, c_m^{(n)}, d_{m_1m_2}^{(n)}$—as well as on the initial data $x_m(0)$—which would guarantee that such a peculiar behavior *always* emerges. Clearly these 2 alternative points of view may *both* have an *applicative* relevance. And clearly the second point of view is the one adopted in this paper.

The findings reported in this paper ascertain which *constraints* on the 60 coefficients $a^{(n)}, b_m^{(n)}, c_m^{(n)}, d_{m_1m_2}^{(n)}$ are *sufficient* to guarantee that *all* solutions $x_n(t)$ of the system of 4 ODEs (1) with (2) be *completely periodic* with a given period $T$; the purpose and scope of this concluding **Section** is to summarize these findings, as reported in the preceding 3 **Sections** of this paper and in its 3 **Appendices**.

The first quite simple set of *constraints* on the 60 coefficients $a^{(n)}$, $b_m^{(n)}$, $c_m^{(n)}$, $d_{m_1 m_2}^{(n)}$ is detailed by the formulas (4), which entail a *very simple* reduction of these *a priori arbitrary* 60 coefficients, to only 45 *different* coefficients.

The remaining restrictions are a bit less simple; they are detailed in **Appendix C**, and they seem to imply that the restrictions on the remaining 45 coefficients are significant but not enormous: from 45 to 34 *arbitrary coefficients*. And this number is further increased—from 34 to 40—if one takes into account the freedom to assign the 6 parameters $f_p$.

So the main result of this paper is to provide fairly explicit formulas allowing to identify the subclass of the system of 4 nonlinearly-coupled ODEs (1) with (2) such that *all* its solutions are *completely periodic* with a *given* period $T$.

It is my hope that these findings shall have some practical use—possibly even in addition to the more specific and detailed results obtained in Ref. [1]—and perhaps as well in other quite different *applicative* contexts in which the system of 4 *nonlinearly-coupled* ODEs (1) with (2) plays a role.

**A final musing**. This paper has been produced in the last days of the year **2024** and presumably shall appear in **arXiv** (the *only* scientific outlet where I now publish my findings) in the first days of the year **2025**; so I end it by offering to its readers, as a **Happy New Year** token, the following (trivial, curious, amusing) *remark*: the *positive integer* **2025** identifying (of course, only in *some* human calendars) this *new year*, is a *rather special* number; as shown by the neatness of its *decomposition* into a product of *primes*: $\mathbf{2025 = 3^4 5^2}$.

**Appendix A**

In this **Appendix A** we display the most obvious consequences of **Ansatz1**. The first 2 eqs. (3c) clearly imply

$$\dot{y}_0(t) = \left[\dot{x}_1(t) - \dot{x}_2(t)\right]/2 ; \qquad (7a)$$

hence, via the 3 eqs. (5), (1) and (3e),

$$Q_2^{(0)}[y_\ell(t)] = \left\{P_2^{(1)}[x_m(t)] - P_2^{(2)}[x_m(t)]\right\} / \left\{2[x_1(0) + x_2(0)]\right\} ; \qquad (7b)$$

hence, via the eqs. (5b), (2) and (4),

$$\alpha^{(0)} + \sum_{\ell_1=0}^{2} \left\{\beta_{\ell_1}^{(0)} y_{\ell_1}(t) + \gamma_{\ell_1}^{(0)} [y_{\ell_1}(t)]^2\right\} + \sum_{\ell_1>\ell_2=0}^{2} \left[\eta_{\ell_1 \ell_2}^{(0)} y_{\ell_1}(t) y_{\ell_2}(t)\right] =$$

$$\left\{a + \sum_{m=1}^{4} \left\{b_m^{(0)} x_m(t) + c_m^{(0)} [x_m(t)]^2\right\} + \right.$$

$$\left. \sum_{m_1>m_2=1}^{4} \left[d_{m_1 m_2}^{(0)} x_{m_1}(t) x_{m_2}(t)\right]\right\} / [x_1(0) + x_2(0)] ; \qquad (7c)$$

which, via the eqs. (3a) and (3b), yield—also via the eqs. (4)—the following 10 *explicit* expressions of the 10 coefficients $\alpha^{(0)}, \beta_\ell^{(0)}, \gamma_\ell^{(0)}$ and $\eta_{\ell_1 \ell_2}^{(0)}$ (with $\ell_1 > \ell_2$) in terms of the 15 coefficients $a^{(0)}$, $b_m^{(0)}$, $c_m^{(0)}$, $d_{m_1 m_2}^{(0)}$ (with $m_1 > m_2$) and the 4 initial values $x_m(0)$:

$$\alpha^{(0)} = \left\{ a^{(1)} + \sum_{m=1}^{4} \left\{ b_m^{(1)} x_m(0) + c_m^{(1)} [x_m(0)]^2 \right\} \right.$$
$$\left. + \sum_{m_1 > m_2 = 1}^{4} \left[ d_{m_1 m_2}^{(1)} x_{m_1}(0) x_{m_2}(0) \right] \right\} / [x_1(0) + x_2(0)] ,$$
(8a)

$$\beta_0^{(0)} = \{ b_1^{(1)} + 2 \left[ c_1^{(1)} x_1(0) - c_2^{(1)} x_2(0) \right]$$
$$+ d_{21}^{(1)} [x_2(0) - x_1(0)] + \left[ d_{31}^{(1)} - d_{32}^{(1)} \right] x_3(0)$$
$$+ \left[ d_{41}^{(1)} - d_{42}^{(1)} \right] x_4(0) \} / [x_1(0) + x_2(0)] ,$$
(8b)

$$\beta_1^{(0)} = [ b_3^{(1)} + 2 c_3^{(1)} x_3(0) + c_1^{(1)} x_1(0)$$
$$+ d_{32}^{(1)} x_2(0) + d_{43}^{(1)} x_4(0) ] / [x_1(0) + x_2(0)] ,$$
(8c)

$$\beta_2^{(0)} = [ b_4^{(1)} + 2 c_4^{(1)} x_4(0) + d_{41}^{(1)} x_1(0)$$
$$+ d_{42}^{(1)} x_2(0) + d_{43}^{(1)} x_3(0) ] / [x_1(0) + x_2(0)] ,$$
(8d)

$$\gamma_0^{(0)} = \left( c_1^{(1)} + c_2^{(1)} - d_{21}^{(1)} \right) / [x_1(0) + x_2(0)] ,$$
(8e)

$$\gamma_1^{(0)} = \left( c_1^{(1)} + c_2^{(1)} - d_{21}^{(1)} \right) / [x_1(0) + x_2(0)] ,$$
(8f)

$$\gamma_2^{(0)} = c_3^{(1)} / [x_1(0) + x_2(0)] ,$$
(8g)

$$\eta_{10}^{(0)} = \left( d_{31}^{(1)} - d_{32}^{(1)} \right) / [x_1(0) + x_2(0)] ,$$
(8h)

$$\eta_{20}^{(0)} = \left( d_{41}^{(1)} - d_{42}^{(1)} \right) / [x_1(0) + x_2(0)] ,$$
(8i)

$$\eta_{21}^{(0)} = d_{43}^{(1)} / [x_1(0) + x_2(0)] .$$
(8j)

While the 2 eqs. (3b) clearly imply

$$y_1(t) = x_3(t) - x_3(0) , \quad y_2(t) = x_4(t) - x_4(0) ,$$
(9a)

hence

$$\dot{y}_1(t) = \dot{x}_3(t) , \quad \dot{y}_2(t) = \dot{x}_4(t) ,$$
(9b)

hence (see eqs. (5) and (1))

$$Q_2^{(s)}[y_\ell] = P_2^{(s+2)}(x_m) / [x_1(0) + x_2(0)] , \quad s = 1, 2 .$$
(10)

Hence, via eqs. (5b) and (2),

$$\alpha^{(s)} + \sum_{\ell=0}^{2} \left\{ y_\ell(t) \left[ \beta_\ell^{(s)} + \gamma_\ell^{(s)} y_\ell(t) \right] \right\} + \sum_{\ell_1 > \ell_2 = 0}^{2} \left[ \eta_{\ell_1 \ell_2}^{(s)} y_{\ell_1}(t) y_{\ell_2}(t) \right] =$$

$$a^{(s+2)} + \left\{ \sum_{m=1}^{4} \left\{ x_m(t) \left[ b_m^{(s+2)} + c_m^{(s+2)} x_m(t) \right] \right\} + \right.$$

$$\left. \sum_{m_1 > m_2 = 1}^{4} \left[ d_{m_1 m_2}^{(s+2)} x_{m_1}(t) x_{m_2}(t) \right] \right\} / [x_1(0) + x_2(0)] \; ; \quad (11)$$

and by replacing the variables $x_m$ in the right-hand sides of these 2 formulas via the eqs. (3a) and (3b), we easily obtain the following expressions of the 20 coefficients $\alpha^{(s)}, \beta_\ell^{(s)}, \gamma_\ell^{(s)}$ and $\eta_{\ell_1 \ell_2}^{(s)}$ (with $\ell_1 > \ell_2$) in terms of the 30 coefficients $a^{(s+2)}$, $b_m^{(s+2)}$, $c_m^{(s+2)}$, $d_{m_1 m_2}^{(s+2)}$ (with $m_1 > m_2$) and the 4 initial values $x_n(0)$:

$$\alpha^{(s)} = \{ a^{(s+2)} + \sum_{m=1}^{4} \left\{ b_m^{(s+2)} x_m(0) + c_m^{(s+2)} [x_m(0)]^2 \right\}$$

$$+ \sum_{m_1 > m_2 = 0}^{4} \left[ d_{m_1 m_2}^{(s+2)} x_{m_1}(0) x_{m_2}(0) \right] \} / [x_1(0) + x_2(0)] \; , \quad (12a)$$

$$\beta_0^{(s)} = \left\{ b_1^{(s+2)} + 2 \sum_{m=1}^{4} \left[ c_m^{(s+2)} x_m(0) \right] + \right.$$

$$\left. \sum_{m_1 > m_2 = 0}^{4} \left\{ d_{m_1 m_2}^{(s+2)} [x_{m1}(0) - x_{m2}(0)] \right\} \right\} / [x_1(0) + x_2(0)] \; , \quad (12b)$$

$$\beta_1^{(s)} = \left[ b_3^{(s+2)} + 2 c_3^{(s+2)} \right] / [x_1(0) + x_2(0)] \; , \quad (12c)$$

$$\beta_2^{(s)} = \left[ b_4^{(s+2)} + 2 c_2^{(s+2)} \right] / [x_1(0) + x_2(0)] \; , \quad (12d)$$

$$\gamma_0^{(s)} = \left[ c_1^{(s+2)} - c_2^{(s+2)} \right] / [x_1(0) + x_2(0)] \; , \quad (12e)$$

$$\gamma_1^{(s)} = c_3^{(s+2)} / [x_1(0) + x_2(0)] \; , \quad (12f)$$

$$\gamma_2^{(s)} = c_3^{(s+2)} / [x_1(0) + x_2(0)] \; , \quad (12g)$$

$$\eta_{10}^{(s)} = \left[ d_{31}^{(s+2)} + d_{32}^{(s+2)} \right] / [x_1(0) + x_2(0)] \; , \quad (12h)$$

$$\eta_{20}^{(s)} = \left[ d_{41}^{(s+2)} - d_{42}^{(s+2)} \right] / [x_1(0) + x_2(0)] \; , \quad (12i)$$

$$\eta_{21}^{(s)} = d_{43}^{(s+2)} / [x_1(0) + x_2(0)] \; . \quad (12j)$$

**Appendix B**

In this **Appendix B** we report some relevant consequences of the additional **Ansatz2**.

By $t$-differentiating the 3 eqs. (6) we get

$$\dot{y}_\ell(t) = \omega \left[ f_{1+2\ell} \cos(\omega t) + f_{2+2\ell} \sin(\omega t) \right] . \tag{13}$$

Hence from the 2 eqs. (5) we get

$$-\omega f_{1+2\ell} \cos(\omega t) - \omega f_{2+2\ell} \sin(\omega t) + \alpha^{(\ell)}$$

$$+ \sum_{\ell_1=0}^{2} \{ \beta^{(\ell)}_{\ell_1} [f_{2+2\ell_1} + f_{1+2\ell_1} \sin(\omega t) - f_{2+2\ell_1} \cos(\omega t)]$$

$$+ \gamma^{(\ell)}_{\ell_1} [f_{2+2\ell_1} + f_{1+2\ell_1} \sin(\omega t) - f_{2+2\ell_1} \cos(\omega t)]^2 \}$$

$$+ \sum_{\ell_1 > \ell_2 = 0}^{2} \{ \eta^{(\ell)}_{\ell_1 \ell_2} [f_{2+2\ell_1} + f_{1+2\ell_1} \sin(\omega t) - f_{2+2\ell_1} \cos(\omega t)] \cdot$$

$$\cdot [f_{2+2\ell_2} + f_{1+2\ell_2} \sin(\omega t) - f_{2+2\ell_2} \cos(\omega t)] \} = 0 . \tag{14}$$

This formula—which must of course hold for *all* values of the independent variable $t$—shall imply (at most) $3 \cdot 5 = 15$ relations—to be satisfied by the $1 + 6 = 7$ parameters $\omega$ and $f_p$—because the $t$-dependence appears only via *linear combinations* with $t$-independent coefficients of only 5 different quantities, namely $1, \sin(\omega t), \cos(\omega t), [\sin(\omega t)]^2, \sin(\omega t)\cos(\omega t)$: indeed, due to the *quadratic* character of the polynomials, no other ($t$-dependent) term shall appear (except for $[\cos(\omega t)]^2$, which however need *not* be included because $[\cos(\omega t)]^2 = 1 - [\sin(\omega t)]^2$).

These 15 formulas are displayed below (each triplet of formulas originating from the factor multiplying one of the 5 quantities $1, \sin(\omega t), \cos(\omega t), \sin(\omega t)^2, \sin(\omega t)\cos(\omega t)$ in the preceding eq. (14)):

$$\alpha^{(\ell)} + \sum_{\ell_1=0}^{2} \left[ \beta^{(\ell)}_{\ell_1} f_{2+2\ell_1} + 2\gamma^{(\ell)}_{\ell_1} (f_{2+2\ell_1})^2 \right] + E^{(\ell)} = 0 , \tag{15a}$$

$$-\omega f_{2+2\ell} + \sum_{\ell_1=0}^{2} \left\{ f_{2+2\ell_1} \left[ \beta^{(\ell)}_{\ell_1} + 2\gamma^{(\ell)}_{\ell_1} f_{1+2\ell_1} \right] \right\}$$

$$+ \sum_{\ell_1 > \ell_2 = 0}^{2} \left[ \eta^{(\ell)}_{\ell_1 \ell_2} (f_{1+2\ell_1} f_{2+2\ell_2} + f_{2+2\ell_1} f_{1+2\ell_2}) \right] = 0 , \tag{15b}$$

$$\omega f_{1+2\ell} + \sum_{\ell_1=0}^{2} \left\{ f_{2+2\ell_1} \left[ \beta^{(\ell)}_{\ell_1} + 2\gamma^{(\ell)}_{\ell_1} f_{1+2\ell_1} \right] \right\} + E^{(\ell)} = 0 , \tag{15c}$$

$$\sum_{\ell_1=0}^{2}\left\{\gamma_{\ell_1}^{(\ell)}\left[(f_{1+2\ell_1})^2-(f_{2+2\ell_1})^2\right]\right\}$$

$$+\sum_{\ell_1>\ell_2=0}^{2}\left[\eta_{\ell_1\ell_2}^{(\ell)}\left(f_{1+2\ell_1}f_{1+2\ell_2}-f_{2+2\ell_1}f_{2+2\ell_2}\right)\right]=0\,,\tag{15d}$$

$$2\sum_{\ell_1=0}^{2}\left[\gamma_{\ell_1}^{(\ell)}f_{1+2\ell_1}f_{2+2\ell_1}\right]+\sum_{\ell_1>\ell_2=0}^{2}\left[\eta_{\ell_1\ell_2}^{(\ell)}\left(f_{1+2\ell_1}f_{2+2\ell_2}+f_{2+2\ell_1}f_{1+2\ell_2}\right)\right]=0\,;\tag{15e}$$

where

$$E^{(\ell)}=2\sum_{\ell_1>\ell_2=0}^{2}\left[\eta_{\ell_1\ell_2}^{(\ell)}f_{2+2\ell_1}f_{2+2\ell_2}\right]\,.\tag{16}$$

Of course these 15 formulas (15) may be either considered to imply 15 (*linear*) *constraints* on the 30 coefficients $\alpha^{(\ell)}$, $\beta_{\ell_1}^{(\ell)}$, $\gamma_{\ell_1}^{(\ell)}$, $\eta_{\ell_1\ell_2}^{(\ell)}$ (with $\ell_1>\ell_2$; for assigned values of the 7 parameters $\omega$ and $f_p$), or alternatively to provide 15 (*nonlinear*) *constraints* on the 7 parameters $\omega$ and $f_p$ (in terms of the 30 coefficients $\alpha^{(\ell)}$, $\beta_{\ell_1}^{(\ell)}$, $\gamma_{\ell_1}^{(\ell)}$, $\eta_{\ell_1\ell_2}^{(\ell)}$): this second set of *constraints* are however *only linear* for the 3 *parameters* $f_p$ with *odd* $p=1,3,5$ (see eqs. (15b) and (15c)), in which case they may of course be *explicitly* solved; for instance for $p=1,3,5$ the *triplet* of eqs. (15c) read as follows:

$$\left[\omega+2\gamma_0^{(0)}f_2\right]\cdot f_1+2\gamma_1^{(0)}f_4\cdot f_3+2\gamma_2^{(0)}f_6\cdot f_5 = -\sum_{\ell_1=0}^{2}\left[f_{2+2\ell_1}\beta_{\ell_1}^{(0)}\right]-E^{(0)}\,,$$

$$2\gamma_0^{(1)}f_2\cdot f_1+\left[\omega+2\gamma_1^{(1)}f_4\right]\cdot f_3+2\gamma_2^{(1)}f_6\cdot f_5 = -\sum_{\ell_1=0}^{2}\left[f_{2+2\ell_1}\beta_{\ell_1}^{(1)}\right]-E^{(1)}\,,$$

$$2\gamma_0^{(2)}f_2\cdot f_1+2\gamma_2^{(2)}f_4\cdot f_3+\left[\omega+2\gamma_2^{(2)}f_6\right]\cdot f_5 = -\sum_{\ell_1=0}^{2}\left[f_{2+2\ell_1}\beta_{\ell_1}^{(2)}\right]-E^{(2)}\,;$$

$$\tag{17a}$$

or, equivalently,

$$\left[\omega\cdot\delta_{0\ell}+2\gamma_0^{(\ell)}f_2\right]\cdot f_1+\left[\omega\cdot\delta_{1\ell}+2\gamma_1^{(\ell)}f_4\right]\cdot f_3+\left[\omega\cdot\delta_{2\ell}+2\gamma_2^{(\ell)}f_6\right]\cdot f_5$$

$$=-\sum_{\ell_1=0}^{2}\left[f_{2+2\ell_1}\beta_{\ell_1}^{(\ell)}\right]-2\sum_{\ell_1>\ell_2=0}^{2}\left[\eta_{\ell_1\ell_2}^{(\ell)}f_{2+2\ell_1}f_{2+2\ell_2}\right]\,,\quad\ell=0,1,2\,;\tag{17b}$$

or equivalently

$$\boldsymbol{\Phi}\cdot\mathbf{f}_{odd}=\boldsymbol{\varphi}\,,\tag{17c}$$

where of course $\boldsymbol{\Phi}$ is the $3 \times 3$ matrix

$$\boldsymbol{\Phi} = \begin{pmatrix} \omega + 2\gamma_0^{(0)} f_2 & 2\gamma_1^{(0)} f_4 & 2\gamma_2^{(0)} f_6 \\ 2\gamma_0^{(1)} f_2 & \omega + 2\gamma_1^{(1)} f_4 & 2\gamma_2^{(1)} f_6 \\ 2\gamma_0^{(2)} f_2 & 2\gamma_2^{(2)} f_4 & \omega + 2\gamma_2^{(2)} f_6 \end{pmatrix} \qquad (17d)$$

and the 3-vectors $\mathbf{f}_{odd}$ and $\boldsymbol{\varphi}$ are defined as follows:

$$\mathbf{f}_{odd} = \begin{pmatrix} f_1 \\ f_3 \\ f_5 \end{pmatrix} , \qquad (17e)$$

$$\boldsymbol{\varphi} = \begin{pmatrix} -\sum_{\ell_1=0}^{2} \left[ f_{2+2\ell_1} \beta_{\ell_1}^{(0)} \right] - 2\sum_{\ell_1 > \ell_2 = 0}^{2} \left[ \eta_{\ell_1 \ell_2}^{(0)} f_{2+2\ell_1} f_{2+2\ell_2} \right] \\ -\sum_{\ell_1=0}^{2} \left[ f_{2+2\ell_1} \beta_{\ell_1}^{(1)} \right] - 2\sum_{\ell_1 > \ell_2 = 0}^{2} \left[ \eta_{\ell_1 \ell_2}^{(1)} f_{2+2\ell_1} f_{2+2\ell_2} \right] \\ -\sum_{\ell_1=0}^{2} \left[ f_{2+2\ell_1} \beta_{\ell_1}^{(2)} \right] - 2\sum_{\ell_1 > \ell_2 = 0}^{2} \left[ \eta_{\ell_1 \ell_2}^{(2)} f_{2+2\ell_1} f_{2+2\ell_2} \right] \end{pmatrix} ; \qquad (17f)$$

so that the 3 *explicit* expressions of the 3 *odd*-numbered parameters $f_1, f_3, f_5$—in terms of the parameter $\omega$, the other 3 *even*-numbered parameters $f_2, f_4, f_4$, and the $3^3 = 27$ coefficients $\beta_{\ell_1}^{(\ell)}, \gamma_{\ell_1}^{(\ell)}, \eta_{\ell_1 \ell_2}^{(\ell)}$ (with $\ell_1 > \ell_2$) read as follows:

$$\mathbf{f}_{odd} = \boldsymbol{\Phi}^{-1} \cdot \boldsymbol{\varphi} . \qquad (17g)$$

But probably these formulas, or analogous ones that the diligent reader might easily write out, are unlikely to be of much use.

**Appendix C**

In this **Appendix C** we ascertain the restrictions on the 45 coefficients $a^{(\nu)}, b_m^{(\nu)}, c_m^{(\nu)}, d_{m_1 m_2}^{(\nu)}$ (with $m_1 > m_2$) implied by the 2 *Ansatzen* introduced above, if one assumes *only* that the 7 parameters $\omega$ and $f_p$ are given, as well as the 4 *initial* data $x_n(0)$; as these are presumably the more relevant assumptions to be *generally* relevant in important *applicative* contexts (for instance, such as those described in Ref. [1]); note that we took already account of the important constraint (4), by replacing the upper index $n$ with the index $\nu$ (and of course the number 60 with 45).

The first step is to ascertain the restrictions on the 30 "Greek" coefficients $\alpha^{(\ell)}, \beta_{\ell_1}^{(\ell)}, \gamma_{\ell_1}^{(\ell)}, \eta_{\ell_1 \ell_2}^{(\ell)}$ (with $\ell_1 > \ell_2$) implied by the $3 \cdot 5 = 15$ eqs. (15) (for any given assignment of the 7 parameters $\omega$ and $f_p$). Since these eqs. (15) are *all linear algebraic* equations, this is a *quite standard* task which may be performed *explicitly* quite easily, except for the difficulty of dealing with the fairly large number of quantities we now like to determine (yes, 15 is usually considered to be a fairly large number: perhaps because we humans have only 10 fingers in our 2 hands, and the 10 extra ones in our feet are of course much less handy; moreover displaying a 15 by 15 matrix is not a quite easy typographical task, and perhaps not quite necessary, so below we will refrain from doing so *explicitly*). Moreover we need to select which 15 "Greek" coefficients we wish to compute (at this stage) in terms of the other (remaining) 15 "Greek" coefficients, while assuming the 7 parameters $\omega$ and $f_p$ to be *arbitrarily assigned* real numbers. We make here a somewhat arbitrary decision in this respect, since our presentation is merely meant to provide an *example* of how to deal with this problematique; in other specific applicative contexts in which the system of Ordinary Differential Equations (ODEs) (1) investigated in this paper might be relevant, other choices might of course be chosen; but the computations would then be quite similar.

So we now identify the following $3+2\cdot2\cdot3 = 15$ dependent variables $\alpha^{(\ell)}$, $\beta_\ell^{(s)}$ $\gamma_\ell^{(s)}$as being those to be computed by solving the 15 eqs. (15), and we directly identify their expressions as solutions of the 15 eqs. (15) by writing

$$\begin{aligned}
\alpha^{(\ell)} &= F_\ell^{(\alpha)}\left(\beta_{\ell_1}^{(0)}, \gamma_{\ell_2}^{(0)}; \omega, f_p\right) , \\
\beta_\ell^{(s)} &= F_{\ell s}^{(\beta)}\left(\beta_{\ell_1}^{(0)}, \gamma_{\ell_2}^{(0)}; \omega, f_p\right) , \\
\gamma_\ell^{(s)} &= F_{\ell s}^{(\gamma)}\left(\beta_{\ell_1}^{(0)}, \gamma_{\ell_2}^{(0)}; \omega, f_p\right) ;
\end{aligned} \qquad (18)$$

we do not display the *explicit* definitions of the 15 functions (of the $2\cdot3+1+6 = 13$ variables $\beta_{\ell_1}^{(0)}, \gamma_{\ell_2}^{(0)}, \omega, f_p$) appearing in the right-hand sides of these 15 eqs. (18), since they are easily computable consequences of the 15 *linear algebraic* eqs. (15) (in close analogy to what was done in **Appendix B**); hence any reader interested in using these findings may quite easily obtain them (up to the possible—but mainly just typographical—difficulty to deal with 15-*components vectors*, and 15-*by*-15 *matrices* as well as their *inverses*).

The next step is to insert these findings (18) in the left-hand side of eq. (8a) (to replace there $\alpha^{(0)}$), and in the left-hand sides of the first 14 eqs. (12) (from (12a) to (12g); these might seem just 7 formulas, but they are actually 14 different equations because $s = 1, 2$); so we have altogether a system of 15 *linear algebraic equations* which allow to compute *explicitly* 15 (*arbitrarily chosen*) of the 30 coefficients $a^{(s+2)}$, $b_m^{(s+2)}$, $c_m^{(s+2)}$ and $d_{m_1 m_2}^{(s+2)}$ (with $m_1 > m_2$); which are then given by *explicit* formulas—which we do not feel the need to display—in terms of the 4 *initial* data $x_m(0)$, as well as the $2+2\cdot2\cdot3 = 14$ coefficients $\alpha^{(s)}$, $\beta_\ell^{(s)}$, $\gamma_\ell^{(s)}$ (already computed above in terms of the 13 quantities $\beta_\ell^{(0)}, \gamma_\ell^{(0)}; \omega, f_p$: see the eqs. (18)). For instance—not to interfere with the following development— let us assume that these 15 (*arbitrarily chosen*) coefficients are all selected out of the $2+2\cdot2\cdot4 = 18$ coefficients $a^{(s+2)}$, $b_m^{(s+2)}$, $c_m^{(s+2)}$ (hence leaving only 3 of these as 18 *freely assignable parameters*).

The only remaining *constraints* to be taken into account are the last 6 eqs. (12h), (12i), (12j) (which again might seem to be just 3, but are in fact 6 because they feature the index $s$); they allow to easily compute 6 of the 10 coefficients $d_{31}^{(s+2)}$, $d_{32}^{(s+2)}$, $d_{41}^{(s+2)}$, $d_{42}^{(s+2)}$, $d_{43}^{(s+2)}$ in terms of the 7 quantities $\eta_{10}^{(s)}$, $\eta_{20}^{(s)}$, $\eta_{21}^{(s)}$ and $x_1(0) + x_2(0)$, and of course of the remaining 4 of the 10 coefficients $d_{31}^{(s+2)}$, $d_{32}^{(s+2)}$, $d_{41}^{(s+2)}$, $d_{42}^{(s+2)}$, $d_{43}^{(s+2)}$. For instance one might write (but this is only one possible option):

$$d_{4\mu}^{(s+2)} = G^{(\mu)}\left(\eta_{10}^{(s)}, \eta_{20}^{(s)}, \eta_{21}^{(s)}; d_{31}^{(s+2)}, d_{32}^{(s+2)}; x_1(0) + x_2(0)\right), \qquad (19)$$

where the 3 functions $G^{(\mu)}\left(\eta_{10}^{(s)}, \eta_{20}^{(s)}, \eta_{21}^{(s)}; d_{31}^{(s+2)}, d_{32}^{(s+2)}; x_1(0) + x_2(0)\right)$—recall that the index $\mu$ only takes the 3 values $\mu = 1, 2, 3$—are of course clearly quite easily obtained from the 6 *linear algebraic* eqs. (12h), (12i), (12j) (but we do not feel the need to actually display these 3 functions, which, again, may be easily computed by anybody interested in using these findings). We thus see that only 4 of the 10 coefficients $d_{31}^{(s+2)}$, $d_{32}^{(s+2)}$, $d_{41}^{(s+2)}$, $d_{42}^{(s+2)}$, $d_{43}^{(s+2)}$ remain as *freely assignable parameters*.

We seem to conclude that of the 45 "Latin" coefficients $a^{(\nu)}, b_m^{(\nu)}, c_m^{(\nu)}, d_{m_1 m_2}^{(\nu)}$ only $15 + 3 + 4 = 22$ are *freely assignable*; while the other 23 are indeed computable in terms of the other 22 *freely assignable* "Latin" coefficients, the 7 parameters $\omega$ and $f_p$, and the 4 *initial data* $x_n(0)$; but in addition also in terms of the $3 + 3 + 3 \cdot 2 = 12$ "Greek" coefficients $\beta_{\ell_1}^{(0)}, \gamma_{\ell_2}^{(0)}, \eta_{10}^{(s)}, \eta_{20}^{(s)}, \eta_{21}^{(s)}$ (see the 45 eqs. (18) and (19)); and since these 12 "Greek" coefficients are themselves *freely assignable* it is reasonable to conclude that the freedom to assign the "Latin" coefficients $a^{(\nu)}, b_m^{(\nu)}, c_m^{(\nu)}, d_{m_1 m_2}^{(\nu)}$ gets significantly increased, from 22 to $22 + 12 = 34$.